\documentclass{ws-ijmpa}

\begin{document}

\begin{center} 
{\Large {\bf Scalar Sector in the Minimal Supersymmetric 3-3-1 model.}}
\end{center}

\begin{center}
M. C. Rodriguez  \\
mcrodriguez@fisica.furg.br \\
{\it Funda\c c\~ao Universidade Federal do Rio Grande-FURG \\
Departamento de F\'\i sica \\
Av. It\'alia, km 8, Campus Carreiros \\
96201-900, Rio Grande, RS \\
Brazil}
\end{center}

\date{\today}

\begin{abstract}
We consider the minimal supersymmetric extension of the 3-3-1 model and we 
study the mass spectra in the scalar sector of this model without the 
anti-sextet. We show that all our lighest scalars are in agreement with the 
experimental limits.
\end{abstract}

PACS number(s): 12.60. Jv, 12.60.Fr.



\section{Introduction}

The Higgs mechanism plays a central role in gauge theories, and still remains 
one of the most indefinite part of the standard model (SM) \cite{sm}. However, 
we know that the SM is not considered as the ultimate theory since neither the 
fundamental parameters, masses and couplings, nor the symmetry pattern are 
predicted. These elements are merely built into the model. 

On another hand, the posibility of a gauge symmetry based on the following 
symmetry $SU(3)_{C} \otimes SU(3)_{L} \otimes U(1)_{N}$ (3-3-1) \cite{ppf,mpp} 
is particularly interesting. The main motivations to study this kind of 
model are:
\begin{enumerate}
\item The family number must be three. This result comes from the fact that the model is anomaly-free only if we 
have equal number of triplets and antitriplets, counting the 
$SU(3)_{c}$ colors, and further more requiring the sum of all fermion charges 
to vanish. However each generation is anomalous, the anomaly 
cancellation occurs for the three, or multiply of three, together and not
generation by generation like in the SM. Therefore triangle anomalies together with asymptotic freedom imply that the number of generations 
must be three and only three. This may provides a first step towards answering the flavor question;
\item It explains why $\sin^{2} \theta_{W}<\frac{1}{4}$ is observed at the 
Z-pole.
This point come from the fact that in the model of Ref.~\cite{ppf} we have that
the $U(1)_N$ and $SU(3)_L$ coupling constants, $g'$ and $g$, respectively,
are related by  
\begin{equation}
t^{2}\equiv \left( \frac{g^{\prime}}{g} \right)^{2}= 
\frac{ \sin^{2} \theta_{W}}{1-4 \sin^{2} \theta_{W}}.
\label{eq2}
\end{equation}
Hence, this 331 model predicts that there exists an energy scale, 
say $\mu$, at which the model loses its perturbative character. 
The value of $\mu$ can be found through the condition   
$\sin^2\theta_W(\mu)=1/4$, and according to recent calculation 
$\mu \approx 4$ TeV \cite{fram,alex};
\item It is the simplest model that includes bileptons of both types: scalar 
and vectors ones. In fact, although there are several models which include doubly charged
scalar fields, not many of them incorporate doubly
charged vector bosons: this is a particularity of the 331 model
of Ref.~\cite{ppf};
\item The model has several sources of CP violation.
In the 331 model~\cite{ppf} we can implement the violation of the CP
symmetry, spontaneously~\cite{laplata1,dumm1} or explicitly~\cite{liung}. 
In models with exotic leptons it is possible to implement soft CP 
violation~\cite{cp3}; 
\item The extra neutral vector boson $Z^\prime$ conserves
flavor in the leptonic but not in the quark sector. The couplings to the leptons
are leptophobic because of the suppression factor $(1-4 \sin^{2} \theta_{W})^{1/2}$ but with 
some quarks there are enhancements because of the factor $(1-4 \sin^{2} \theta_{W})^{-1/2}$
~\cite{dumm2}.
\end{enumerate}
Recently, we have proposed the supersymmetric extensions of the 331 models 
\cite{first,331susy1,331susy2}. Since it is possible to define the $R$-parity symmetry in both models, the phenomenology 
with $R$-parity conserved has similar features to that of the 
$R$-conserving MSSM: the supersymmetric particles are pair-produced and the
lightest neutralino is the lightest supersymmetric particle (LSP). 
However, there are differences between this kind of models and the MSSM with or without $R$-parity breaking.

In the case in the model of Ref.~\cite{331susy1}, there are doubly charged scalar and vector fields. Hence, 
we have doubly charged charginos which are mixtures of the superpartners of the $U$-vector boson with the 
doubly charged scalars \cite{mcr}. This implies new interactions that are not present in the 
MSSM, for instance: $\tilde{\chi}^{--}\tilde{\chi}^0U^{++}$, $\tilde{\chi}^{-}\tilde{\chi}^{-}U^{++}$, 
$\tilde{l}^{-}l^{-}\tilde{\chi}^{++}$ where $\tilde{\chi}^{++}$ denotes any doubly charged chargino. 
Moreover, in the chargino production, besides the usual mechanism, we have additional contributions coming 
from the $U$-bilepton in the s-channel. Due to this fact we have an enhancement in the cross section of production 
of these particles in $e^{-}e^{-}$ collisors, such as the NLC~\cite{mcr}. 

We will also have the singly charged charginos 
and neutralinos, as in the MSSM, where there are processes like $\tilde{l}^{-}l^{+}\tilde{\chi}^{0}$, 
$\tilde{\nu_{L}}l^{-}\tilde{\chi}^{+}$, with $\tilde{l}$ denoting any slepton; $\tilde{\chi}^{-}$ denotes singly charged 
chargino and $\tilde{\nu}_L$ denotes any sneutrino. The only difference is that in the MSSM there are five neutralinos
while in our model there are eight neutralinos.

In the case of model in Ref.~\cite{331susy2} there is no double charged chargino. The mechanism of production 
of charginos and neutralinos are the same as in the MSSM, but in the case of $R$-parity conservation there are fifteen 
neutralinos and six singly charged chargino in this model.

The scalar sector of the minimal 331 model \cite{ppf} was studied on 
\cite{tonasse}. While the scalar sector of the 331 model with right handed 
neutrinos is given at \cite{l97}. Recently it was presented the scalar sector 
of the supersymmetric 331 model with right-handed neutrinos \cite{huong}. On 
this work we will analyse the scalar sector of the minimal supersymmetric 
331 model.

This paper is organized as follows. In Sec. \ref{psusy1} we review
the minimal supersymmetric 331 model. The construction of the
supersymmetric scalar potential is discussed in Sec.
\ref{sec:sp}, while in Sec. \ref{sec:constraint} we give the constraint 
equations. In \ref{sec:analyses} the numerical
analysis are given. Finally, the last section is
devoted to our conclusions.

\section{Minimal Supersymmetric 331 Model}
\label{psusy1}

In the nonsupersymmetric 331 model~\cite{ppf} the fermionic representation
content is as follows: left-handed leptons 
$L_{aL}=(\nu_{a},l_{a},l^{c}_{a})_{L}\sim({\bf1},{\bf3},0)$, 
$a=e,\mu,\tau$; left-handed quarks 
$Q_{\alpha L}=(d_{\alpha},u_{\alpha},j_{\alpha})_{L}\sim({\bf3},{\bf3}^*,-1/3)$
, $\alpha=1,2$, $Q_{3L}=(u_{3},d_{3},J)_{L}\sim({\bf3},{\bf3},2/3)$; and in 
the right-handed components we have $u^{c}_{iL}\sim({\bf3}^{*},{\bf1},-2/3),
d^{c}_{iL}\sim({\bf3}^{3},{\bf1},1/3),\,i=1,2,3$, and the exotic quarks 
$j^{c}_{\alpha L}\sim({\bf3}^{*},{\bf1},4/3),J^{c}_{L}\sim({\bf3}^{*},{\bf1},
-5/3)$. While in the supersymmetric version of this model we have to add their 
supersymmetric partners $\tilde{L}_{aL}$, $\tilde{Q}_{\alpha L}$, 
$\tilde{Q}_{3L}$, $\tilde{u}^{c}_{iL}$, $\tilde{d}^{c}_{iL}$, 
$\tilde{j}^{c}_{\alpha L}$ and $\tilde{J}^{c}_{L}$ \cite{331susy1}.

The minimal scalar representation content is formed by three scalar triplets: 
$\eta\sim({\bf1},{\bf3},0)=(\eta^{0},\eta^{-}_{1},\eta^{+}_{2})^T$;
$\rho\sim({\bf1},{\bf3},+1)=(\rho^{+}, \rho^{0}, \rho^{++})^T$ and 
$\chi\sim({\bf1},{\bf3},-1)=(\chi^{-},\chi^{--},\chi^{0})^T$, and their 
supersymmetric partner $\tilde{\eta}$, $\tilde{\rho}$ and $\tilde{\chi}$. 
However, we have to introduce the followings extras scalars 
$\eta^{\prime}$, $\rho^{\prime}$, $\chi^{\prime}$ and their higgsinos 
$\tilde{\eta}^{\prime}$, $\tilde{\rho}^{\prime}$ and $\tilde{\chi}^{\prime}$.

In the nonsupersymmetric 331 model to give arbitrary mass to the leptons we 
have to introduce one scalar 
antisextet $S\sim({\bf1},{\bf6}^{*},0)$.We can avoid the introduction of the 
antisextet by adding a charged lepton transforming as a singlet. 
Notwithstanding, here we will omit both the antisextet 
and the exotic lepton, because we have showed in \cite{lepmass} that in this 
model the $R$-violating interactions give the correct masses to $e,\mu$ and 
$\tau$. 

The complete set of fields in the minimal supersymmetric 331 model
(MSUSY331) has been given in \cite{331susy1,mcr}, see these articles to the 
complete Lagrangian of the model. On this article we will write only the 
lagrangian necessary to construct the mass spectra in the scalar sector.

The gauge sector's lagrangian is wrtting as
\begin{eqnarray}
{\cal L}^{gauge} 
&=&  \frac{1}{4} \int  d^{2}\theta\;Tr[W_{C}W_{C}]+ \frac{1}{4} 
\int  d^{2}\theta\;Tr[W_{L}W_{L}]+
\frac{1}{4} \int  d^{2}\theta W^{ \prime}W^{ \prime} \nonumber \\ &+&  
\frac{1}{4} \int  d^{2}\bar{\theta}\;Tr[\bar{W}_{C}\bar{W}_{C}]+ \frac{1}{4} 
\int  d^{2}\bar{\theta}\;Tr[\bar{W}_{L}\bar{W}_{L}]+
\frac{1}{4} \int  d^{2}\bar{\theta} \bar{W}^{ \prime}\bar{W}^{ \prime}\,\ , 
\nonumber \\
\label{gaugm1}
\end{eqnarray}
while the lagrangian of the scalar sector is written as
\begin{eqnarray}
{\cal L}^{scalar} 
&=& \int d^{4}\theta\;\left[\,\hat{ \bar{ \eta}}e^{2g\hat{V}} \hat{ \eta} + 
\hat{ \bar{ \rho}}e^{ \left( 2g\hat{V}+g^{\prime}\hat{V}^{\prime} \right)} \hat{ \rho} +
\hat{ \bar{ \chi}}e^{ \left( 2g\hat{V}-g^{\prime}\hat{V}^{\prime} \right)} \hat{ \chi} \right. \nonumber \\
&+& \left.\,\hat{ \bar{ \eta}}^{\prime}e^{2g\hat{ \bar{V}}} \hat{ \eta}^{\prime} + 
\hat{ \bar{ \rho}}^{\prime}e^{ \left( 2g\hat{ \bar{V}}-g^{\prime}\hat{V}^{\prime} \right)} \hat{ \rho}^{\prime} +
\hat{ \bar{ \chi}}^{\prime}e^{ \left( 2g\hat{ \bar{V}}+g^{\prime}\hat{V}^{\prime} \right)} \hat{ \chi}^{\prime} \right]
+ \int d^{2}\theta W+ \int d^{2}\bar{ \theta}\overline{W} \,\ , \nonumber \\
\label{escm1}
\end{eqnarray}
here $g$ and $g^{\prime}$ are the gauge coupling constants of $SU(3)$ and 
$U(1)$ respectivelly and $W$ is the superpotential of the model given by 
$W=W_{2}+W_{3}$.

The superpotential is written explicity as
\begin{eqnarray}
W_{2}&=&\mu_{0a}\hat{L}_{aL} \hat{ \eta}^{\prime}+ 
\mu_{ \eta} \hat{ \eta} \hat{ \eta}^{\prime}+
 \mu_{ \rho} \hat{ \rho} \hat{ \rho}^{\prime}+ 
\mu_{ \chi} \hat{ \chi} \hat{ \chi}^{\prime}, \nonumber \\
W_{3}&=& \lambda_{1abc} \epsilon \hat{L}_{aL} \hat{L}_{bL} \hat{L}_{cL}+
\lambda_{2ab} \epsilon \hat{L}_{aL} \hat{L}_{bL} \hat{ \eta}+ 
\lambda_{3a} \epsilon 
\hat{L}_{aL} \hat{\chi} \hat{\rho}+
f_{1} \epsilon \hat{ \rho} \hat{ \chi} \hat{ \eta}+
f^{\prime}_{1}\epsilon \hat{ \rho}^{\prime}\hat{ \chi}^{\prime}\hat{ \eta}^{\prime} \nonumber \\
&+&
\kappa_{1\alpha i} \hat{Q}_{\alpha L} \hat{\rho} \hat{u}^{c}_{iL} +  
\kappa_{2\alpha i} \hat{Q}_{\alpha L} \hat{\eta} \hat{d}^{c}_{iL}+
\kappa_{3\alpha \beta} \hat{Q}_{\alpha L} \hat{\chi} \hat{j}^{c}_{\beta L} \nonumber \\
&+& 
\kappa_{4\alpha ai} \hat{Q}_{\alpha L} \hat{L}_{aL} \hat{d}^{c}_{iL}+ 
\kappa_{5i} \hat{Q}_{3L} \hat{\eta}^{\prime} \hat{u}^{c}_{iL}+
\kappa_{6i} \hat{Q}_{3L} \hat{\rho}^{\prime} \hat{d}^{c}_{iL}+
\kappa_{7} \hat{Q}_{3L} \hat{\chi}^{\prime} \hat{J}^{c}_{L} \nonumber \\
&+&
\xi_{1ijk} \hat{d}^{c}_{iL} \hat{d}^{c}_{jL} \hat{u}^{c}_{kL}+
\xi_{2ij \beta} \hat{u}^{c}_{iL} \hat{u}^{c}_{jL} \hat{j}^{c}_{\beta L}+
\xi_{3i \beta} \hat{d}^{c}_{iL} \hat{J}^{c}_{L} \hat{j}^{c}_{\beta L}. 
\label{sp3m1}
\end{eqnarray}
The coefficients $\mu_{0}, \mu_{\eta}, \mu_{\rho}$ and $\mu_{\chi}$ have mass 
dimension, while all the coefficients in $W_{3}$ are dimensionless.

To get the scalar potential of our model we have to eliminate the auxiliarly 
fields $F$ and $D$ that appear in our model. We are going to pick up the $F$ 
and $D$- terms, from Eqs.(\ref{gaugm1},\ref{escm1},\ref{sp3m1}), we get
\begin{eqnarray}
{\cal L}^{gauge}_{D}&=&\frac{1}{2}D^{a}D^{a}+ \frac{1}{2}DD \,\ , \nonumber \\
{\cal L}^{scalar}_{F}&=& \vert F_{\eta} \vert^2+ \vert F_{\rho} \vert^2+ 
\vert F_{\chi} \vert^2+  
\vert F_{\eta^{\prime}} \vert^2+ 
\vert F_{\rho^{\prime}} \vert^2+ 
\vert F_{\chi^{\prime}} \vert^2, \nonumber \\
{\cal L}^{scalar}_{D}&=& \frac{g}{2} \left[ \bar{\eta}\lambda^a\eta+ 
\bar{\rho}\lambda^a\rho+ \bar{\chi}\lambda^a\chi- 
\bar{\eta}^{\prime}\lambda^{* a}\eta^{\prime}- 
\bar{\rho}^{\prime}\lambda^{* a}\rho^{\prime}- 
\bar{\chi}^{\prime}\lambda^{* a}\chi^{\prime} \right] D^{a} \nonumber \\
&+& \frac{g^{ \prime}}{2} \left[  
\bar{\rho}\rho- \bar{\chi}\chi- \bar{\rho}^{\prime}\rho^{\prime}+ 
\bar{\chi}^{\prime}\chi^{\prime} \right]D, \nonumber \\
{\cal L}^{W2}_{F}&=& \frac{\mu_{ \eta}}{2}( \eta F_{\eta^{\prime}}+ 
\eta^{\prime} F_{ \eta}+ \bar{\eta} \bar{F}_{\eta^{\prime}}+ 
\bar{\eta^{\prime}} \bar{F}_{ \eta})+ 
\frac{\mu_{ \rho}}{2}( \rho F_{\rho^{\prime}}+ \rho^{\prime} F_{ \rho}+
\bar{\rho} \bar{F}_{\rho^{\prime}}+ \bar{\rho^{\prime}} \bar{F}_{ \rho}) 
\nonumber \\
&+& \frac{\mu_{ \chi}}{2}( \chi F_{\chi^{\prime}}+ \chi^{\prime} F_{ \chi}+
\bar{\chi} \bar{F}_{\chi^{\prime}}+ \bar{\chi^{\prime}} \bar{F}_{ \chi}), 
\nonumber \\
{\cal L}^{W3}_{F}&=& \frac{1}{3} \left[
f_{1} \epsilon (F_{ \rho} \chi \eta+ \rho F_{ \chi} \eta+ \rho \chi F_{ \eta}+
\bar{F}_{ \rho} \bar{\chi} \bar{\eta}+ \bar{\rho} \bar{F}_{ \chi} \bar{\eta}+ 
\bar{\rho} \bar{\chi} \bar{F}_{ \eta} ) \right. \nonumber \\
&+& \left. f^{\prime}_{1} \epsilon (
F_{ \rho^{\prime}} \chi^{\prime} \eta^{\prime}+ 
\rho^{\prime} F_{ \chi^{\prime}} \eta^{\prime}+ 
\rho^{\prime} \chi^{\prime} F_{ \eta^{\prime}}+
\bar{F}_{ \rho^{\prime}} \bar{\chi^{\prime}} \bar{\eta^{\prime}}+ 
\bar{\rho^{\prime}} \bar{F}_{ \chi^{\prime}} \bar{\eta^{\prime}}+ 
\bar{\rho^{\prime}} \bar{\chi^{\prime}} \bar{F}_{ \eta^{\prime}}) \right] . 
\nonumber \\
\end{eqnarray}

From the equation described above we can construct
\begin{eqnarray}
{\cal L}_{F}&=&{\cal L}^{scalar}_{F}+{\cal L}^{W2}_{F}+{\cal L}^{W3}_{F} 
\nonumber \\
&=&\vert F_{\eta} \vert^2+ \vert F_{\rho} \vert^2+ 
\vert F_{\chi} \vert^2+  
\vert F_{\eta^{\prime}} \vert^2+ 
\vert F_{\rho^{\prime}} \vert^2+ 
\vert F_{\chi^{\prime}} \vert^2 \nonumber \\
&+& \frac{1}{2} \left[ \mu_{ \eta} ( \eta F_{\eta^{\prime}}+ 
\eta^{\prime} F_{ \eta}+ \bar{\eta} \bar{F}_{\eta^{\prime}}+ 
\bar{\eta^{\prime}} \bar{F}_{ \eta})+ 
\mu_{ \rho} ( \rho F_{\rho^{\prime}}+ \rho^{\prime} F_{ \rho}+
\bar{\rho} \bar{F}_{\rho^{\prime}}+ \bar{\rho^{\prime}} \bar{F}_{ \rho})  
\right. \nonumber \\
&+& \left. \mu_{ \chi} ( \chi F_{\chi^{\prime}}+ 
\chi^{\prime} F_{ \chi}+
\bar{\chi} \bar{F}_{\chi^{\prime}}+ \bar{\chi^{\prime}} \bar{F}_{ \chi}) 
\right] + \frac{1}{3} \left[ f_{1} \epsilon (F_{ \rho} \chi \eta+ \rho F_{ \chi} \eta+ \rho \chi F_{ \eta} \right. \nonumber \\
&+& \left.
\bar{F}_{ \rho} \bar{\chi} \bar{\eta}+ \bar{\rho} \bar{F}_{ \chi} \bar{\eta}+ 
\bar{\rho} \bar{\chi} \bar{F}_{ \eta} ) + f^{\prime}_{1} \epsilon (
F_{ \rho^{\prime}} \chi^{\prime} \eta^{\prime}+ 
\rho^{\prime} F_{ \chi^{\prime}} \eta^{\prime}+ 
\rho^{\prime} \chi^{\prime} F_{ \eta^{\prime}} \right. \nonumber \\
&+& \left.
\bar{F}_{ \rho^{\prime}} \bar{\chi^{\prime}} \bar{\eta^{\prime}}+ 
\bar{\rho^{\prime}} \bar{F}_{ \chi^{\prime}} \bar{\eta^{\prime}}+ 
\bar{\rho^{\prime}} \bar{\chi^{\prime}} \bar{F}_{ \eta^{\prime}})
 \right] \,\ , \nonumber \\
{\cal L}_{D}&=&{\cal L}^{gauge}_{D}+{\cal L}^{scalar}_{D} \nonumber \\
&=& \frac{1}{2}D^{a}D^{a}+ \frac{1}{2}DD 
+ \frac{g}{2} \left[ \bar{\eta}\lambda^a\eta+ 
\bar{\rho}\lambda^a\rho+ \bar{\chi}\lambda^a\chi- 
\bar{\eta}^{\prime}\lambda^{* a}\eta^{\prime}- 
\bar{\rho}^{\prime}\lambda^{* a}\rho^{\prime} \right. \nonumber \\
&-& \left. 
\bar{\chi}^{\prime}\lambda^{* a}\chi^{\prime} \right] D^{a}+ 
\frac{g^{ \prime}}{2} \left[  
\bar{\rho}\rho- \bar{\chi}\chi- \bar{\rho}^{\prime}\rho^{\prime}+ 
\bar{\chi}^{\prime}\chi^{\prime} \right]D.
\label{auxiliarm1}
\end{eqnarray}

We will now show that these fields can be eliminated through the 
Euler-Lagrange equations
\begin{eqnarray}
\frac{\partial {\cal L}}{\partial \phi}- \partial_{m} 
\frac{\partial {\cal L}}{\partial (\partial_{m} \phi)}=0 \,\ ,  
\label{Euler-Lagrange Equation}
\end{eqnarray}
where 
$\phi = \eta , \rho , \chi , \eta^{\prime}, \rho^{\prime}, \chi^{\prime}$. 
Formally auxiliary fields are defined 
as fields having no kintetic terms. Thus, this definition immediately yields 
that the Euler-Lagrange equations for auxiliary fields simplify to 
$\frac{\partial {\cal L}}{\partial \phi}=0$.

Applying these simplified equations to various auxiliary $F$-fields yields 
the following relations
\begin{eqnarray}
\bar{F}_{\eta}&=&- \left( \frac{\mu_{\eta}}{2}\eta^{\prime}+ 
\frac{f_{1}}{3} \epsilon \rho \chi \right) \,\ ;
\,\ F_{\eta}=- \left( \frac{\mu_{\eta}}{2}\bar{\eta^{\prime}}+ 
\frac{f_{1}}{3} \epsilon \bar{\rho} \bar{\chi} \right) \,\ , \nonumber \\
\bar{F}_{\rho}&=&- \left( \frac{\mu_{\rho}}{2}\rho^{\prime}+ 
\frac{f_{1}}{3} \epsilon \chi \eta \right) \,\ ;
\,\ F_{\rho}=- \left( \frac{\mu_{\rho}}{2}\bar{\rho^{\prime}}+ 
\frac{f_{1}}{3} \epsilon \bar{\chi} \bar{\eta} \right) \,\ , \nonumber \\
\bar{F}_{\chi}&=&- \left( \frac{\mu_{\chi}}{2}\chi^{\prime}+ 
\frac{f_{1}}{3} \epsilon \rho \eta \right) \,\ ;
\,\ F_{\chi}=- \left( \frac{\mu_{\chi}}{2}\bar{\chi^{\prime}}+ 
\frac{f_{1}}{3} \epsilon \bar{\rho} \bar{\eta} \right) \,\ , \nonumber \\
\bar{F}_{\eta^{\prime}}&=&- \left( \frac{\mu_{\eta}}{2}\eta + 
\frac{f^{\prime}_{1}}{3} \epsilon \rho^{\prime}\chi^{\prime} 
\right) \,\ ;
\,\  F_{\eta^{\prime}}=- \left( \frac{\mu_{\eta}}{2}\bar{\eta}+ 
\frac{f^{\prime}_{1}}{3} \epsilon \bar{\rho^{\prime}}\bar{\chi^{\prime}} 
\right) \,\ , \nonumber \\
\bar{F}_{\rho^{\prime}}&=&- \left( \frac{\mu_{\rho}}{2}\rho + 
\frac{f^{\prime}_{1}}{3} \epsilon \chi^{\prime}\eta^{\prime} 
\right) \,\ ;
\,\ F_{\rho^{\prime}}=- \left( \frac{\mu_{\rho}}{2}\bar{\rho}+ 
\frac{f^{\prime}_{1}}{3} \epsilon \bar{\chi^{\prime}}\bar{\eta^{\prime}} 
\right) \,\ , \nonumber \\
\bar{F}_{\chi^{\prime}}&=&- \left( \frac{\mu_{\chi}}{2}\chi + 
\frac{f^{\prime}_{1}}{3} \epsilon \rho^{\prime} \eta^{\prime} 
\right) \,\ ;
\,\  F_{\chi^{\prime}}=- \left( \frac{\mu_{\chi}}{2}\bar{\chi}+ 
\frac{f^{\prime}_{1}}{3} \epsilon \bar{\rho^{\prime}}\bar{\eta^{\prime}} 
\right) \,\ ,
\end{eqnarray}
using these equations, we can rewrite Eq.(\ref{auxiliarm1}) as
\begin{equation}
{\cal L}_{F}=- \left(
\vert F_{\eta} \vert^2+ \vert F_{\rho} \vert^2+ \vert F_{\chi} \vert^2+  
\vert F_{\eta^{\prime}} \vert^2+ \vert F_{\rho^{\prime}} \vert^2+ 
\vert F_{\chi^{\prime}} \vert^2 \right) .
\end{equation}
If we perform the same program to $D$-fields we get
\begin{eqnarray}
D^{a}&=&- \frac{g}{2} \left[ \bar{\eta}\lambda^a\eta+ 
\bar{\rho}\lambda^a\rho+ \bar{\chi}\lambda^a\chi- 
\bar{\eta}^{\prime}\lambda^{* a}\eta^{\prime}- 
\bar{\rho}^{\prime}\lambda^{* a}\rho^{\prime}- 
\bar{\chi}^{\prime}\lambda^{* a}\chi^{\prime} \right] \,\ , \nonumber \\
D&=&- \frac{g^{ \prime}}{2} \left[  
\bar{\rho}\rho- \bar{\chi}\chi- \bar{\rho}^{\prime}\rho^{\prime}+ 
\bar{\chi}^{\prime}\chi^{\prime} \right] \,\ ,
\end{eqnarray}
According with Eq.(\ref{auxiliarm1})
\begin{equation}
{\cal L}_{D}=- \frac{1}{2} \left( D^{a}D^{a}+DD \right) \,\ .
\end{equation}

\section{The scalar potential}
\label{sec:sp}

The pattern of the symmetry breaking of the model is given by the following 
scheme
\begin{eqnarray}
&\mbox{MSUSY331}&
\stackrel{{\cal L}_{soft}}{\longmapsto}
\mbox{SU(3)}_C\ \otimes \ \mbox{SU(3)}_L\otimes \mbox{U(1)}_N
\stackrel{\langle\chi\rangle \langle \chi^{\prime}\rangle}{\longmapsto}
\mbox{SU(3)}_C \ \otimes \ \mbox{SU(2)}_L\otimes
\mbox{U(1)}_Y \nonumber \\
&\stackrel{\langle\rho,\eta, \rho^{\prime}\eta^{\prime}\rangle}{\longmapsto}&
\mbox{SU(3)}_C \ \otimes \ \mbox{U(1)}_Q
\end{eqnarray}
From this pattern of the symmetry breaking comes the constraint \cite{mcr}
\begin{equation} 
V^2_\eta+V^2_\rho=(246\;{\rm GeV})^2
\label{wmasslimite}
\end{equation} 
coming from $M_W$, where, we have defined $V^2_\eta=v^2_\eta+v^{\prime2}_\eta$ 
and $V^2_\rho= v^2_\rho+v^{\prime2}_\rho$.

The scalar potential is written as
\begin{equation}
V_{MSUSY331}=V_{D}+V_{F}+V_{\mbox{soft}}
\label{ep1}
\end{equation}
where

\begin{eqnarray}
V_{D}&=&-{\cal L}_{D}=\frac{1}{2}\left(D^{a}D^{a}+DD\right)\nonumber \\ &=&
\frac{g^{\prime 2}}{2}(\bar{\rho}\rho-\bar{\rho^{\prime}}\rho^{\prime}
-\bar{\chi}\chi+\bar{\chi^{\prime}}\chi^{\prime})^{2}+
\frac{g^{2}}{8}\sum_{i,j}\left(\bar{\eta}_{i}\lambda^{a}_{ij}\eta_{j}
+\bar{\rho}_{i}\lambda^{a}_{ij}\rho_{j}
+\bar{\chi}_{i}\lambda^{a}_{ij}\chi_{j}\right.
\nonumber \\ &-&
\left.\bar{\eta^{\prime}}_{i}\lambda^{*a}_{ij}\eta^{\prime}_{j} 
-\bar{\rho^{\prime}}_{i}\lambda^{*a}_{ij}\rho^{\prime}_{j}
-\bar{\chi^{\prime}}_{i}\lambda^{*a}_{ij}\chi^{\prime}_{j} \right)^{2}, 
\nonumber \\
V_{F}&=&-{\cal L}_{F}=\sum_{m}\bar{F}_{m}F_{m}\nonumber \\ &=&
\sum_{i,j,k}\left[
\left\vert \frac{\mu_{\eta}}{2} \eta^{\prime}_{i}+\frac{f_{1}}{3}
\epsilon_{ijk}\rho_{j}\chi_{k} \right\vert^{2}+
\left\vert \frac{\mu_{\rho}}{2}\rho^{\prime}_{i}+\frac{f_{1}}{3}
\epsilon_{ijk}\chi_{j}\eta_{k} \right\vert^{2}+.
\left\vert \frac{\mu_{\chi}}{2}\chi^{\prime}_{i}+\frac{f_{1}}{3}
\epsilon_{ijk}\rho_{j}\eta_{k} \right\vert^{2} \right. \nonumber \\
&+& \left.
\left\vert \frac{\mu_{\eta}}{2} \eta_{i}+\frac{f^{\prime}_{1}}{3}
\epsilon_{ijk}\rho^{\prime}_{j}\chi^{\prime}_{k} \right\vert^{2}+
\left\vert \frac{\mu_{\rho}}{2}\rho_{i}+\frac{f^{\prime}_{1}}{3}
\epsilon_{ijk}\chi^{\prime}_{j}\eta^{\prime}_{k} \right\vert^{2}+
\left\vert \frac{\mu_{\chi}}{2}\chi_{i}+\frac{f^{\prime}_{1}}{3}
\epsilon_{ijk}\rho^{\prime}_{j}\eta^{\prime}_{k} \right\vert^{2} \right], 
\nonumber \\
V_{soft}&=&-{\cal L}^{\mbox{scalar}}_{soft}\nonumber \\ &=&
m^{2}_{\eta}\bar{\eta}\eta+
m^{2}_{\rho}\bar{\rho}\rho+
m^{2}_{\chi}\bar{\chi}\chi+
m^{2}_{\eta^{\prime}}\bar{\eta^{\prime}}\eta^{\prime}+
m^{2}_{\rho^{\prime}}\bar{\rho^{\prime}}\rho^{\prime}+
m^{2}_{\chi^{\prime}}\bar{\chi^{\prime}}\chi^{\prime} \nonumber \\ &+&
[k_{1}\epsilon\,\rho\chi\eta+
k^{\prime}_{1}\epsilon\,\rho^{\prime}\chi^{\prime}\eta^{\prime}+ H.c.],
\label{ess}
\end{eqnarray}
where, $m^{2}_{\eta}, m^{2}_{\rho}, m^{2}_{\chi}, m^{2}_{\eta^{\prime}}, 
m^{2}_{\rho^{\prime}}, m^{2}_{\chi^{\prime}},k_{1}$ and $k^{\prime}_{1}$ 
have mass dimension.

All the six neutral scalar components 
$\eta^0,\rho^0,\chi^0,\eta^{\prime0},\rho^{\prime0},
\chi^{\prime0}$
gain non-zero vacuum expectation values. Making a shift in 
the neutral scalars as
\begin{eqnarray}
< \eta > &=& 
      \left( \begin{array}{c}
v_{\eta}+H_{\eta}+iF_{\eta}
\\ 
                  0 \\
                  0          \end{array} \right) 
\,\ , \,\
< \eta^{\prime} > = 
      \left( \begin{array}{c}
v_{\eta^{\prime}}+H_{\eta^{\prime}}+ iF_{\eta^{\prime}}\\ 
                  0 \\
                  0          \end{array} \right) 
\,\ ,  \nonumber \\
< \rho > &=& 
      \left( \begin{array}{c} 0 \\
v_{\rho}+H_{\rho}+iF_{\rho} \\
                  0          \end{array} \right)
\,\ , \,\ 
< \rho^{\prime} > = 
      \left( \begin{array}{c} 0 \\ 
v_{\rho^{\prime}}+H_{\rho^{\prime}}+iF_{\rho^{\prime}}\\
                  0          \end{array} \right) \,\ ,\nonumber \\
< \chi > &=& 
      \left( \begin{array}{c} 0 \\ 
                  0 \\
v_{\chi}+H_{\chi}+iF_{\chi} \end{array} \right)
\,\ , \,\
< \chi^{\prime} > = 
      \left( \begin{array}{c} 0 \\ 
                  0 \\
v_{\chi^{\prime}}+H_{\chi^{\prime}}+iF_{\chi^{\prime}}\end{array} \right) \,\ .
\label{develop}
\end{eqnarray}

\section{Constraint Equations}
\label{sec:constraint}

Here in this section we give the constraint equations, due to the requirement 
the potential to reach a minimum at the chosen VEV's. We get this equation 
requiring that in the shifted potential the linear terms in fields must 
be absent
\begin{eqnarray}
&&0=
\frac{g^2}{12}(2v^2_\eta-2v^{\prime2}_\eta-v^2_\rho+v^{\prime2}_\rho-
v^2_\chi+v^{\prime2}_\chi)
+m^2_\eta+\frac{1}{4}\mu^2_\eta+\frac{f^2_1}{18}(v^2_\rho+v^2_\chi)
\nonumber \\ &+&\frac{k_1}{\sqrt2}\frac{v_\rho v_\chi}{v_\eta}+
\frac{1}{6\sqrt2}(f_1\frac{\mu_\rho v^\prime_\rho v_\chi}{v_{\eta}}+f_1
\frac{\mu_\chi v_\rho v^\prime_\chi}{v_{\eta}}+f^\prime_1
\frac{\mu_\eta v^\prime_\rho v^\prime_\chi}{v_{\eta}}), \nonumber \\
&&0=
\frac{g^2}{12}(-v^2_\eta+v^{\prime2}_\eta+2v^2_\rho-
2v^{\prime2}_{\rho}-v^2_\chi+v^{\prime2}_\chi)+
\frac{g^{\prime2}}{2}(v^2_\rho-v^{\prime2}_\rho-
v^2_\chi+v^{\prime2}_\chi)\nonumber \\ &+&
\frac{f^2_1}{18}(v^2_\eta+v^2_\chi)+m^2_\rho+\frac{1}{4}\mu^2_\rho+
\frac{k_1}{\sqrt2}\frac{v_\eta v_\chi}{v_\rho}+
\frac{f_1}{6\sqrt2 v_\rho}(\frac{\mu_\eta v^\prime_\eta v_\chi}{v_{\rho}}+
\frac{\mu_\chi v_\eta v^\prime_\chi}{v_{\rho}}) \nonumber \\
&-& \frac{f^{\prime}_{1}}{6 \sqrt{2}} 
\frac{\mu_{\rho}v^{\prime}_{\eta}v^{\prime}_{\chi}}{v_{\rho}} \nonumber \\
&&0=
\frac{g^2}{12}(-v^2_\eta+v^{\prime2}_\eta-v^2_\rho+
v^{\prime2}_\rho+v^2_\chi-
2v^{\prime2}_\chi)+\frac{g^{\prime2}}{2}(v^2_\chi-v^{\prime2}_\chi-v^2_\rho
+v^{\prime2}_\rho)\nonumber \\ &+&
\frac{f^2_1}{18}(v^2_\eta+v^2_\rho)+m^2_\chi+\frac{1}{4}\mu^2_\chi
-\frac{1}{6\sqrt2}(f_1\frac{\mu_\rho v_\eta v^\prime_\rho}{v_{\chi}}-f_1
\frac{\mu_\eta v^\prime_\eta v_\rho}{v_{\chi}} +f^\prime_1
\frac{\mu_\chi v^\prime_\eta v^\prime_\rho}{v_{\chi}}) 
\nonumber \\
&+&\frac{k_1}{\sqrt2}\frac{v_\eta v_\rho}{v_\chi}, \nonumber \\
&&0=
\frac{g^2}{12}(-2v^2_\eta+v^{\prime2}_\eta
+v^2_\rho-v^{\prime2}_\rho+v^2_\chi
-v^{\prime2}_\chi)+m^2_{\eta^\prime}+\frac{1}{4}\mu^2_\eta+
\frac{f^2_1}{18}(v^{\prime2}_\rho+
v^{\prime2}_\chi)\nonumber \\ &+&
\frac{k^\prime_1}{\sqrt2}\frac{v^\prime_\rho v^\prime_\chi}{v^\prime_\eta}
+\frac{1}{6\sqrt2}(f_1\frac{\mu_\eta v_\rho v_\chi}{v^{\prime}_{\eta}}-
f^\prime_1\frac{\mu_\chi v^\prime_\rho v_\chi}{v^{\prime}_{\eta}}), 
\nonumber \\
&&0=
\frac{g^2}{12}(v^2_\eta-v^{\prime2}_\eta-2v^2_\rho+2v^{\prime2}_\rho+v^2_\chi-
v^{\prime2}_\chi)+\frac{g^{\prime2}}{2}(-v^2_\rho+v^{\prime2}_\rho+v^2_\chi
-v^{\prime2}_\chi)\nonumber \\ &+&
\frac{f^\prime_1}{18}(v^{\prime2}_\eta+v^{\prime2}_\chi)+
m^2_{\rho^\prime}+\frac{1}{4}\mu^2_\rho+
\frac{1}{6\sqrt2}(f_1\frac{\mu_\rho v_\eta v_\chi}{v^{\prime}_{\rho}}-
f^\prime_1\frac{\mu_\chi v^\prime_\eta v_\chi}{v^{\prime}_{\rho}}+
f^\prime_1\frac{\mu_\eta v_\eta v^\prime_\chi}{v^{\prime}_{\rho}}) \nonumber \\
&+& \frac{k^{\prime}_{1}}{\sqrt{2}}
\frac{v^\prime_\eta v^\prime_\chi}{v^\prime_\rho}, 
\nonumber \\
&&0=
\frac{g^2}{12}(v^2_\eta-v^{\prime2}_\eta+v^2_\rho-v^{\prime2}_\rho-v^2_\chi
+v^{\prime2}_\chi)+\frac{g^{\prime2}}{2}(v^2_\rho-v^{\prime2}_\rho -v^2_\chi
+v^{\prime2}_\chi)\nonumber \\ &+&
\frac{f^{\prime2}_1}{18}(v^{\prime2}_\eta+v^2_\rho)+
m^2_{\chi^\prime}+\frac{1}{4}\mu^2_\chi+
\frac{1}{6\sqrt2}(f^\prime_1
\frac{\mu_\eta v_\eta v^\prime_\rho}{v^{\prime}_{\chi}}-
f^\prime_1\frac{\mu_\rho v^\prime_\eta v_\rho}{v^{\prime}_{\chi}}+
f_1\frac{\mu_\chi v_\eta v_\rho}{v^{\prime}_{\chi}}) \nonumber \\
&+& \frac{k^\prime_1}{\sqrt2}
\frac{v^\prime_\eta v^\prime_\rho}{v^\prime_\chi}. 
\label{constraintequationsinmodel1}
\end{eqnarray}

The mass matrices, thus, can be calculated, using
\begin{equation}
M_{ij}^2=\frac{\partial ^2V_{MSUSY331}}{\partial \phi _i\partial \phi _j}
\label{calculomassamodelo1}
\end{equation}
evaluated at the chosen minimum, where $\phi_i$ are the scalars of our 
model described above.

For the sake of simplicity, here we assume that 
vacuum expectation values (VEVs) are real. This means that
the CP violation through the scalar exchange is not considered
in this work. In literature, a real part $H$ is called  CP-even scalar or 
{\it scalar}, and an imaginary one  $F$  -  CP-odd scalar or 
{\it pseudoscalar} field. In this paper we call them  scalar and pseudoscalar, 
respectively.

\section{Mass Spectrum general case}
\label{sec:analyses}

We will use below the following set of parameters in the scalar potential:
\begin{equation}
f_1=1.2,\quad f^\prime_1=10^{-6},\quad{\rm (dimensionless)}
\label{fs}
\end{equation}
and
\begin{equation}
-k_1=k^\prime_1=10,\  
-\mu_\eta=\mu_\rho= \mu_\chi=1000, \quad \mbox{(in GeV)},
\label{ks}
\end{equation}
we also used the constraint in Eq.(\ref{wmasslimite}). Assuming that 
$v_\eta=20$, $v_\chi=1000$, $v^\prime_\eta=v^\prime_\rho=v^\prime_\chi=1$ in 
GeV, the value of $v_\rho$ is fixed by the constraint in 
Eq.(\ref{wmasslimite}).

\begin{figure}[ht]
\begin{center}
\vglue -0.009cm
\centerline{\epsfig{file=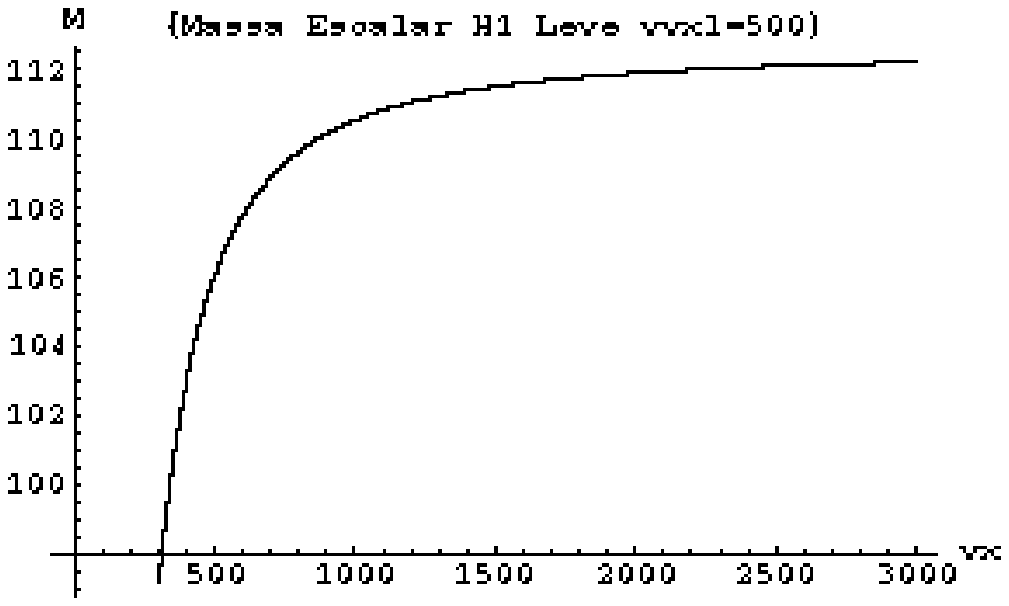,width=0.7\textwidth,angle=0}} 
\end{center}
\caption{The eigenvalues of $m_{H^{0}_{1}}$ as function of $v_{\chi}$.}
\label{fig1}
\end{figure}

\begin{figure}[ht]
\begin{center}
\vglue -0.009cm
\centerline{\epsfig{file=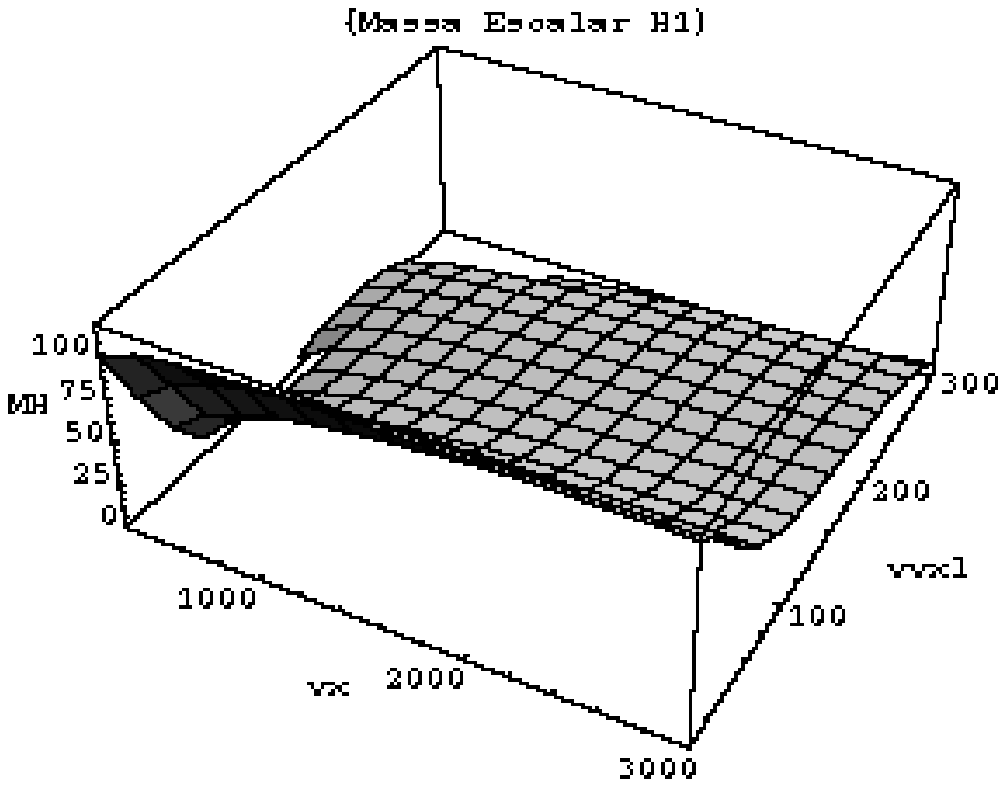,width=0.7\textwidth,angle=0}}
\end{center}
\caption{$m_{H^{0}_{1}} \times v_{\chi} \times v_{\chi^{\prime}}$.}
\label{fig2}
\end{figure}

Diagonalizing the matrices we got the mass eigenstates. On the case of the 
neutral real scalar,Eq.(\ref{matscam1}), we got in (GeV):
\begin{eqnarray}
m_{H^{0}_{1}}&=&110.5 \,\ , m_{H^{0}_{2}}=249 \,\ , m_{H^{0}_{3}}=701.3 \,\ , 
\nonumber \\
m_{H^{0}_{4}}&=&1220 \,\ ,  m_{H^{0}_{5}}=1683.9 \,\ , 
m_{H^{0}_{6}}=4951.7 \,\ ,
\end{eqnarray}
in fig.(\ref{fig1}) we show the behavior of the lightest scalar $H^{0}_{1}$ as 
a function of $v_{\chi}$, although in fig.(\ref{fig2}) we have done the same 
analyses as function of $v_{\chi}$ and $v^{\prime}_{\chi}$. The current $95\%$ 
CL mass bound on the lightest scalar at MSSM is $89.8$GeV \cite{pdg}.

\begin{figure}[ht]
\begin{center}
\vglue -0.009cm
\centerline{\epsfig{file=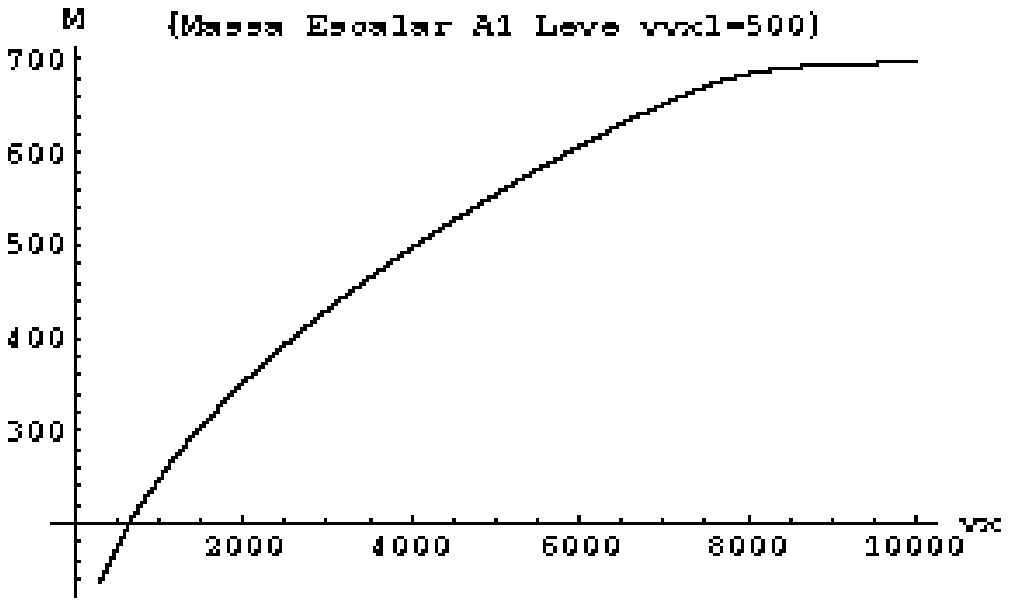,width=0.7\textwidth,angle=0}}
\end{center}
\caption{The eigenvalues of $m_{A_{1}}$ as function of $v_{\chi}$.}
\label{fig3}
\end{figure}

\begin{figure}[ht]
\begin{center}
\vglue -0.009cm
\centerline{\epsfig{file=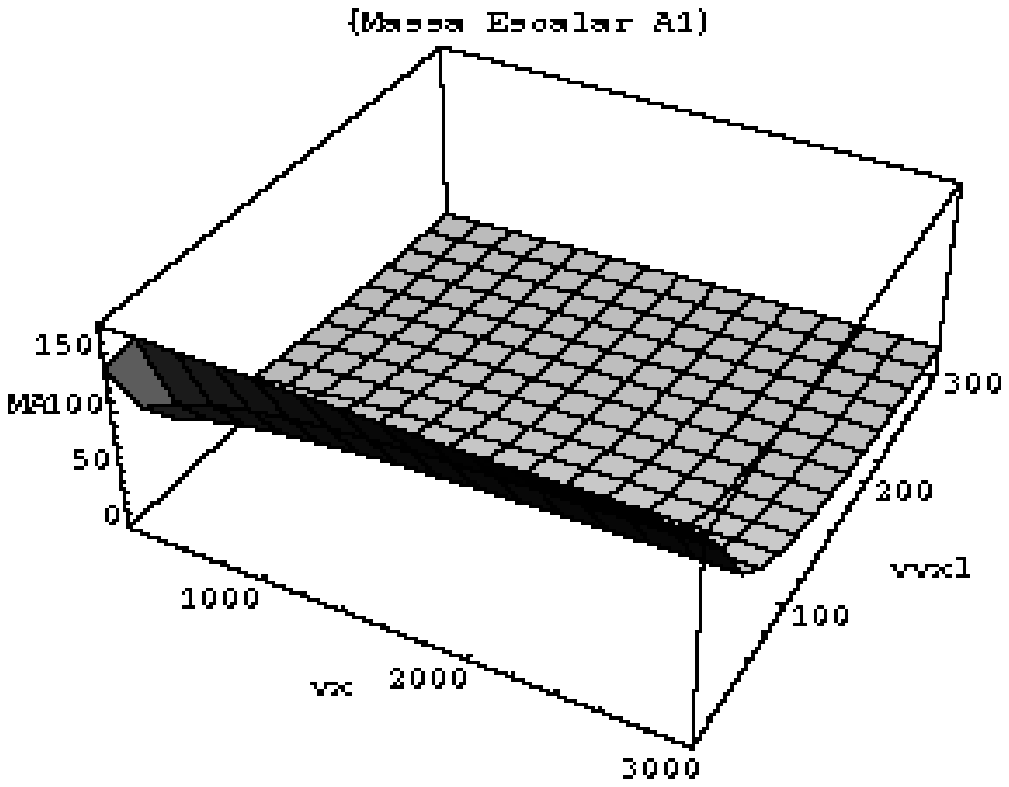,width=0.7\textwidth,angle=0}}
\end{center}
\caption{$m_{A_{1}} \times v_{\chi} \times v^{\prime}_{\chi}$.}
\label{fig4}
\end{figure}

To the neutral imaginary,Eq.(\ref{matpseum1}), we got the following values 
in (GeV)
\begin{eqnarray}
m_{A_{1}}&=&248.6 \,\ , m_{A_{2}}=701.3 \,\ , \nonumber \\
m_{A_{3}}&=&1683.9 \,\ , m_{A_{4}}=4951.7 \,\ .     
\end{eqnarray}
in fig.(\ref{fig3}), we show the behavior of the lightest pseudoscalar 
$A_{1}$ as a function of $v_{\chi}$, although in fig.(\ref{fig4}) we have done 
the same analyses as function of $v_{\chi}$ and $v^{\prime}_{\chi}$. The 
current $95\%$ CL mass bound on the lightest pseudoscalar at MSSM is 
$90.$GeV \cite{pdg}.

\begin{figure}[ht]
\begin{center}
\vglue -0.009cm
\centerline{\epsfig{file=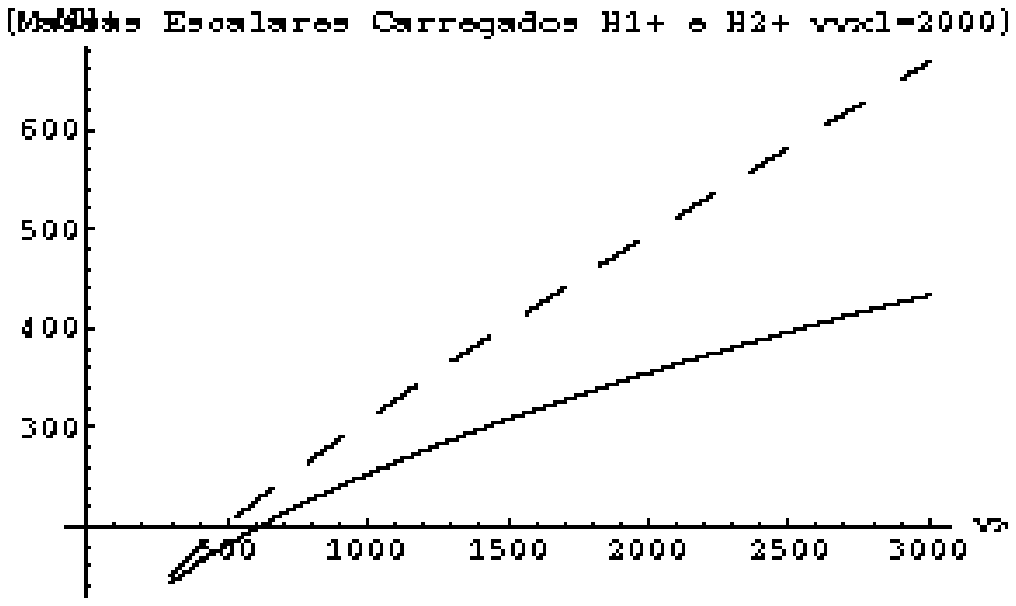,width=0.7\textwidth,angle=0}}
\end{center}
\caption{The eigenvalues of $m_{H^{+}_{1}}$(solid line), $m_{H^{+}_{4}}$
(dashed line),  as function of $v_{\chi}$.}
\label{fig5}
\end{figure}

\begin{figure}[ht]
\begin{center}
\vglue -0.009cm
\centerline{\epsfig{file=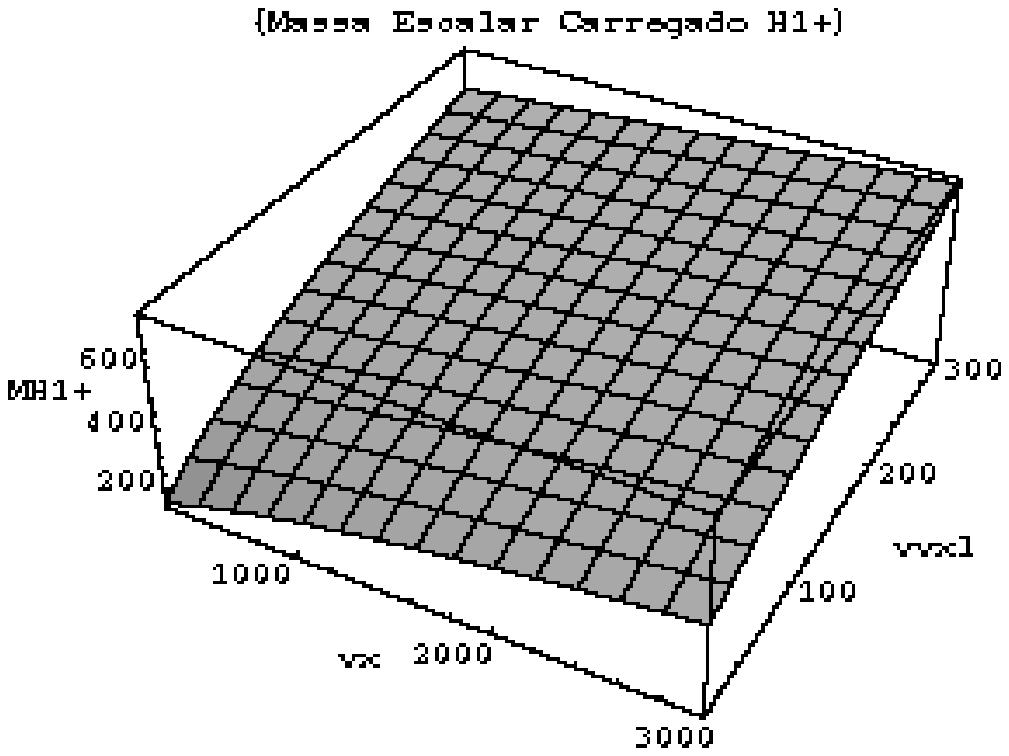,width=0.7\textwidth,angle=0}}
\end{center}
\caption{$m_{H^{+}_{1}} \times v_{\chi} \times v^{\prime}_{\chi}$.}
\label{fig6}
\end{figure}

On the single charged sector considerating the first matrix, 
Eq.(\ref{mixeta1rho}), the physical states have the following masses in (GeV) 
\begin{equation}
m_{H^{+}_{1}}=253 \,\ , m_{H^{+}_{2}}=1684.9 \,\ , m_{H^{+}_{3}}=4951.3
\end{equation}
in fig.(\ref{fig5}) we show the behavior of the eigenvalues of 
$m_{H^{+}_{1}}$(solid line), $m_{H^{+}_{4}}$ (dashed line),  as function of 
$v_{\chi}$ although in fig.(\ref{fig6}) we see the 
dependence of $m_{H^{+}_{1}}$ in terms of $v_{\chi}$ and $v^{\prime}_{\chi}$.

\begin{figure}[ht]
\begin{center}
\vglue -0.009cm
\centerline{\epsfig{file=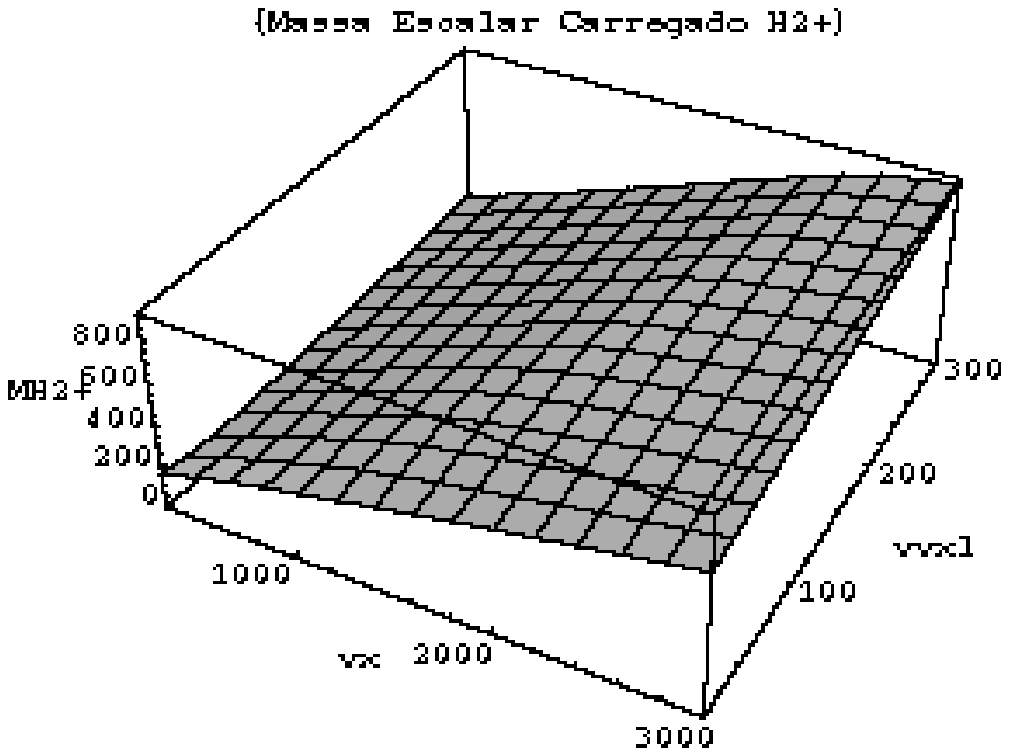,width=0.7\textwidth,angle=0}}
\end{center}
\caption{$m_{H^{+}_{4}} \times v_{\chi} \times v^{\prime}_{\chi}$.}
\label{fig7}
\end{figure}

While in the second matrix, Eq.(\ref{mixeta2chi}), we get 
\begin{equation}
m_{H^{+}_{4}}=308.4 \,\ , m_{H^{+}_{5}}=773.4 \,\ , m_{H^{+}_{6}}=4940.8
\end{equation}
in fig.(\ref{fig7}), we see the dependence of $m_{H^{+}_{4}}$ in terms of 
$v_{\chi}$ and $v^{\prime}_{\chi}$. The current $95\%$ CL mass bound on the 
lightest charged scalar at MSSM is $79.3$GeV \cite{pdg}.

\begin{figure}[ht]
\begin{center}
\vglue -0.009cm
\centerline{\epsfig{file=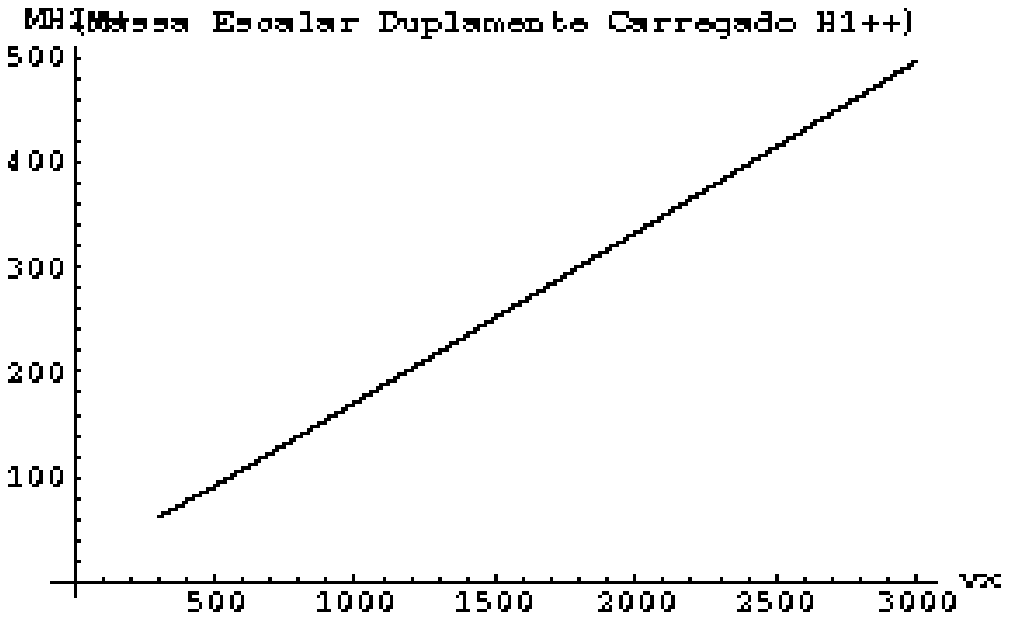,width=0.7\textwidth,angle=0}}
\end{center}
\caption{The eigenvalues of $m_{H^{++}_{1}}$ as function of $v_{\chi}$.}
\label{fig8}
\end{figure}

\begin{figure}[ht]
\begin{center}
\vglue -0.009cm
\centerline{\epsfig{file=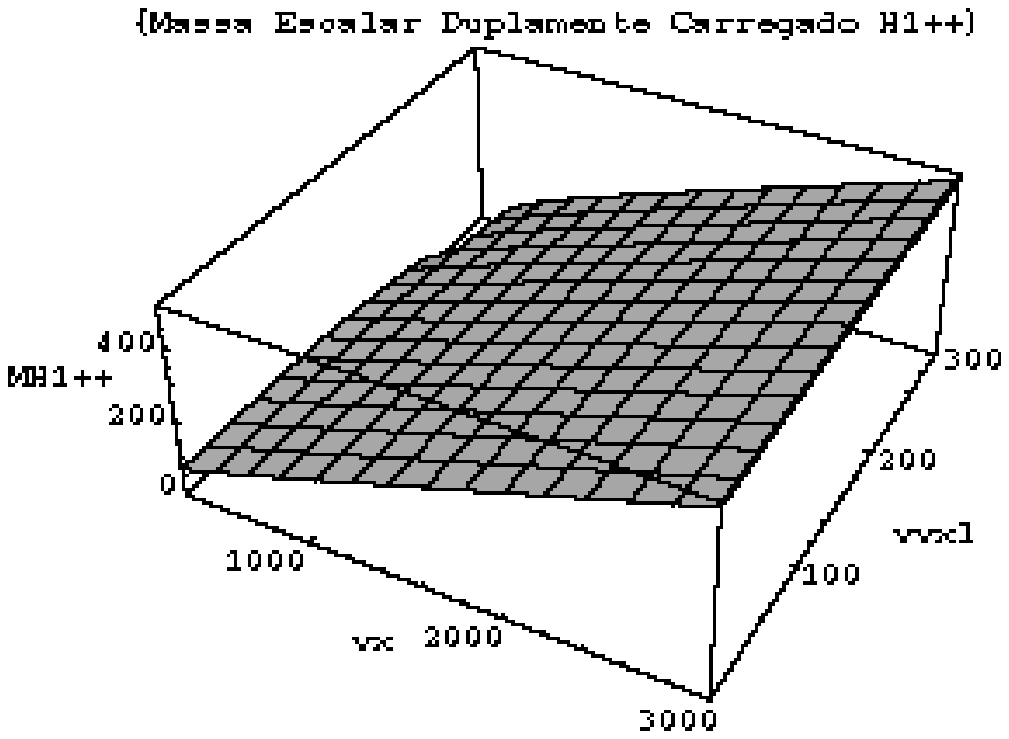,width=0.7\textwidth,angle=0}}
\end{center}
\caption{$m_{H^{++}_{1}} \times v_{\chi} \times v_{\chi^{\prime}}$. 
}
\label{fig9}
\end{figure}

While in the double charged sector, Eq.(\ref{matdcharm1}), there are
\begin{equation}
m_{H^{++}_{1}}=170.7 \,\ , m_{H^{++}_{2}}=770.6 \,\ , 
m_{H^{++}_{3}}=1650.7 \,\ .
\end{equation} 
in fig.(\ref{fig8}) we show the behaviour of $H^{++}_{1}$ as function of 
$v_{\chi}$, although in fig.(\ref{fig9}), we see the dependence of 
$m_{H^{++}_{1}}$ in terms of $v_{\chi}$ and $v^{\prime}_{\chi}$. The current 
$95\%$ CL mass bound on the lightest doubly-charged scalar is 
$95$GeV and $100$GeV were obtained for left-right symmetric models (the exact 
limits depend on the leptons flavors) \cite{pdg}.

\section{Conclusions}
\label{sec:conclusion}

On this article we constructed all the spectrum from the scalar
sector of the minimal supersymmetric 3-3-1 model, and we also show that all 
our lighest scalars are in agreement with the experimental limits, as 
discussed above.

We want also to recall the attention that our results are in agreement with 
those presented on \cite{331susy1,indianos}.

\section{Graphics}




\appendix

\section{Scalar in MSUSY331.}

Calculating Eq.(\ref{ep1}) with the help of 
Eqs.(\ref{develop},\ref{constraintequationsinmodel1}) and 
using as base the following set of 
scalars 
$H_{\eta},H_{\rho},H_{\chi},H_{\eta^{\prime}},H_{\rho^{\prime}},
H_{\chi^{\prime}}$, we get following mass matrix to the Real Neutral 
Fields
\begin{eqnarray}
{\bf M_{11}}&=& \frac{g^2v^2_\eta}{3}+ \frac{1}{6\sqrt{2}v_\eta}
(f_1v^\prime_\rho \mu_\rho v_\chi-6k_1v_\rho v_\chi-f^\prime_1\mu_\eta 
v^\prime_\rho v^\prime_\chi+f_1\mu_\chi v_\rho v^\prime_\chi),
\nonumber \\
{\bf M_{12}}&=&-\frac{g^2v_\eta
v_\rho}{6}+\frac{1}{9\sqrt2}(\sqrt{2}f^2_1 v_\eta v_\rho+9k_1 v_\chi-
\frac{3}{2}\mu_\chi v^\prime_\chi), \nonumber \\
{\bf M_{13}}&=&-\frac{g^2v_\eta
v_\rho}{6}+\frac{1}{9\sqrt2}(9k_1v_\rho- 
\frac{3}{2}f_1\mu_\rho v^\prime_\rho+\sqrt{2}f^2_1v_\eta v_\chi)\nonumber \\
{\bf M_{14}}&=&-\frac{g^2v_\eta v^\prime_\eta}{3},\;
{\bf M_{15}}=\frac{g^2 v_\eta v^\prime_\rho}{6}-
\frac{1}{6\sqrt2}(\mu_\rho v_\chi-\mu_\eta v^\prime_\chi),\nonumber \\ 
{\bf M_{16}}&=&\frac{g^2 v_\eta v^\prime_\chi}{6}+
\frac{1}{6\sqrt2}(f^\prime_1 \mu_\eta v^\prime_\rho-f_1\mu_\chi v_\rho),
\nonumber \\
{\bf M_{22}}&=& \left( \frac{g^2}{3}+g^{\prime 2} \right) v^2_\rho-
\frac{v_\chi}{6\sqrt2 v_\rho}(6k_1v_\eta+ f_1\mu_\eta v^\prime_\eta) 
\nonumber \\
&+& \frac{v^\prime_\chi}{6\sqrt2 v_\rho}(f^\prime_1\mu_\rho
v^\prime_\eta+f_1\mu_\chi v_\eta),
\nonumber \\
{\bf M_{23}}&=&- \left( \frac{g^2}{6}+g^{\prime2} \right) v_\rho v_\chi
+\frac{1}{6\sqrt2}(6k_1v_\eta+f^\prime_1 \mu_\eta v^\prime_\eta)
+ \frac{v_\rho v_\chi}{9}f^2_1,\; \nonumber \\
{\bf M_{24}}&=& \frac{g^2}{6}v^\prime_\eta
v_\rho+\frac{1}{6\sqrt2}(f_1\mu_\eta v_\chi-f^\prime_1\mu_\rho v^\prime_\chi) 
\,\ ,
{\bf M_{25}}=-(\frac{g^2}{3}+g^{\prime2})\frac{v_\rho v^\prime_\rho}{3},
\; \nonumber \\
{\bf M_{26}}&=&\left( \frac{g^2}{6}+g^{\prime2} \right) v_\rho v^\prime_\chi 
-\frac{\mu_\rho}{6\sqrt2}f^\prime_1 v^\prime_\eta -
\frac{\mu_\chi}{6\sqrt2}f_1v_\eta \nonumber \\
{\bf M_{33}}&=& \left( \frac{g^2}{3}+g^{\prime2} \right) v^2_\chi-
\frac{1}{6\sqrt2 v_\chi}
(6k_1v_\eta v_\rho+f_1\mu_\eta v^\prime_\eta
v_\rho-f_1\mu_\rho v_\eta v^\prime_\rho-f^\prime_1\mu_\chi v^\prime_\eta
v^\prime_\rho), \; \nonumber \\
{\bf M_{34}}&=&\frac{g^2}{6}v^\prime_\eta
v_\chi+\frac{1}{6\sqrt2}(f_1\mu_\rho v_\rho-f^\prime_1\mu_\chi v^\prime_\rho),
\nonumber \\ 
{\bf M_{35}}&=& \left( \frac{g^2}{6}+g^{\prime2} \right) v^\prime_\rho v_\chi-
\frac{1}{6\sqrt2}(f_1\mu_\rho v_\eta+f^\prime_1\mu_\chi v^\prime_\eta),
\nonumber \\ 
{\bf M_{36}}&=& - \left( \frac{g^2}{3}+g^{\prime2} \right) v_\chi 
v^\prime_\chi, \nonumber \\ 
{\bf M_{44}}&=&\frac{g^2}{3}v^{\prime2}_\eta+\frac{1}{6\sqrt2 v^\prime_\eta}
(f^\prime_1\mu_\chi v^\prime_\rho v_\chi-f_1\mu_\rho v_\rho
v_\eta-6k^\prime_1 v^\prime_\rho
v^\prime_\chi+f^\prime_1\mu_\rho v_\rho v^\prime_\chi), \nonumber \\
{\bf M_{45}}&=&-\frac{g^2}{6}v^\prime_\eta
v^\prime_\rho+\frac{1}{9\sqrt2}(\sqrt2f^{\prime2}_1v^\prime_\eta
v^\prime_\rho- \frac{3}{2}f^\prime_1\mu_\chi v_\chi+
9k^\prime_1 v^\prime_\chi),\nonumber \\ 
{\bf M_{46}}&=&-\frac{g^2}{6}v^\prime_\eta v^\prime_\chi+
\frac{1}{9\sqrt2}(9k^\prime_1 v^\prime_\rho-\frac{3}{2}f^\prime_1\mu_\rho
v_\rho+\sqrt2 f^{\prime2}_1 v^\prime_\eta v^\prime_\chi), \nonumber \\
{\bf M_{55}}&=& \left( \frac{g^2}{3}+g^{\prime2} \right) v^{\prime2}_\rho+
\frac{1}{6\sqrt2 v^\prime_\rho}(f_1\mu_\rho v_\eta v_\chi+
f^\prime_1\mu_\chi v^\prime_\eta v_\chi -6k^\prime_1v^\prime_\eta 
v^\prime_\chi -f^\prime_1\mu_\eta v_\eta v^\prime_\chi) 
\nonumber \\
{\bf M_{56}}&=&- \left( \frac{g^2}{6}+g^{\prime2} \right) v^\prime_\rho
v^\prime_\chi+\frac{1}{12\sqrt2}(12k^\prime_1v^\prime_\eta+
2f^\prime_1\mu_\eta v_\eta +\frac{3}{\sqrt2}f^{\prime2}_1v^\prime_\rho 
v^\prime_\chi) \nonumber \\ 
{\bf M_{66}}&=& \left( \frac{g^2}{3}+g^{\prime2} \right) v^{\prime2}_\chi-
\frac{1}{6\sqrt2
v^\prime_\chi}(6k^\prime_1v^\prime_\eta v^\prime_\rho+f^\prime_1\mu_\eta 
v_\eta v^\prime_\rho-f^\prime_1\mu_\rho
v^\prime_\eta v_\rho-f_1\mu_\chi v_\eta v_\rho). 
\nonumber \\
\label{matscam1}
\end{eqnarray}

This matrix has no Goldstone bosons and six mass eigenstates, which we 
denote as $H^{0}_{1},H^{0}_{2},H^{0}_{3},H^{0}_{4},H^{0}_{5},H^{0}_{6}$.

\section{Pseudoscalar in MSUSY331.}

On this case using the base given by 
$F_{\eta},F_{\rho},F_{\chi},F_{\eta^{\prime}},F_{\rho^{\prime}},
F_{\chi^{\prime}}$, the mass matrix, with the help of 
Eq.(\ref{constraintequationsinmodel1}), is given by:
\begin{eqnarray}
{\bf {\cal M}_{11}}&=&\frac{1}{6\sqrt2 v_\eta}(
f_1\mu_\rho v^\prime_\rho v^\prime_\chi -6k_1v_\rho v_\chi),\;
{\bf {\cal M}_{12}}=\frac{1}{6\sqrt2}(f_1\mu_\chi-6f_1 v_\chi),\nonumber \\  
{\bf {\cal M}_{13}}&=&\frac{1}{6\sqrt2}(f_1\mu_\rho v^\prime_\rho-6k_1v_\rho),
\;
{\bf {\cal M}_{14}}=0,\;
{\bf {\cal M}_{15}}=\frac{1}{6\sqrt2}(f_1\mu_\eta
v^\prime_\chi-f^\prime_1\mu_\rho v_\chi), \nonumber \\
{\bf {\cal M}_{16}}&=&\frac{1}{6\sqrt2}(f^\prime_1\mu_\eta v^\prime_\rho 
-f_1\mu_\chi v_\rho), \nonumber \\
{\bf {\cal M}_{22}}&=&\frac{1}{6\sqrt2 v_\rho}(-6k_1 v_\eta
v_\chi-f_1\mu_\eta v^\prime_\eta v_\chi+f^\prime_1\mu_\rho
v^\prime_\eta v^\prime_\chi+f_1\mu_\chi v_\eta
v^\prime_\chi),\nonumber \\
{\bf {\cal M}_{23}}&=&- \frac{1}{6\sqrt2}(6k_1v_\eta+f_1\mu_\eta 
v^\prime_\eta),\nonumber \\
{\bf {\cal M}_{24}}&=&
\frac{1}{6\sqrt2}(f_1\mu_\eta v_\chi
-f^\prime_1\mu_\rho v^\prime_\chi),\;{\cal M}_{25}=0,\nonumber \\
{\bf {\cal M}_{26}}&=&- \frac{1}{6\sqrt2}(f^\prime_1\mu_\rho v^\prime_\eta
+f_1\mu_\chi v_\eta), \nonumber \\
{\bf {\cal M}_{33}}&=&\frac{1}{6\sqrt2 v_\chi}(-6f_1 k_1v_\eta
v_\rho-f_1\mu_\eta v^\prime_\eta v_\rho+f_1\mu_\rho v_\eta 
v^\prime_\rho +
f^\prime_1\mu_\chi v^\prime_\eta v^\prime_\rho),\nonumber \\
{\bf {\cal M}_{34}}&=&\frac{1}{\sqrt2}(f_1\mu_\eta v_\rho-f^\prime_1\mu_\chi
v^\prime_\chi),\; {\bf {\cal M}_{35}}=-
\frac{1}{6\sqrt2}(f_1\mu_\rho v_\eta+f^\prime_1\mu_\chi v^\prime_\eta), 
\nonumber \\
{\bf {\cal M}_{36}}&=&0,\nonumber \\
{\bf {\cal M}_{44}}&=&\frac{1}{6\sqrt2 v^\prime_\eta}
(-f_1\mu_\eta v_\rho v_\chi+f^\prime_1
\mu_\chi v^\prime_\rho v_\chi-6k^\prime_1 v^\prime_\rho
v^\prime_\chi+f^\prime_1\mu_\rho v_\rho v^\prime_\chi+f^\prime_1\mu_\rho 
v_\rho v^\prime_\chi),\nonumber \\
{\bf {\cal M}_{45}}&=&0,\;\; 
{\bf {\cal M}_{46}}=\frac{1}{6\sqrt2}(-6k^\prime_1 
v^\prime_\rho+f^\prime_1\mu_\rho v_\rho), \nonumber \\
{\bf {\cal M}_{55}}&=&\frac{1}{6\sqrt2 v^\prime_\rho}
(f_1\mu_\rho v_\eta v_\chi
+f^\prime_1\mu_\chi v^\prime_\eta v_\chi-6k^\prime_1 v^\prime_\eta
v^\prime_\chi-f^\prime_1\mu_\eta v_\eta v^\prime_\chi),\nonumber \\ 
{\bf {\cal M}_{56}}&=&- \frac{1}{6\sqrt2}(6k^\prime_1 v^\prime_\eta+
f^\prime_1\mu_\eta v_\eta), \, \
{\bf {\cal M}_{66}}=\frac{1}{6\sqrt2 v^\prime_\chi}
(f^\prime_1\mu_\rho v^\prime_\eta v_\rho -
f^\prime_1\mu_\eta v_\eta v^\prime_\rho).
\label{matpseum1}
\end{eqnarray}

This mass matrix has two Goldstone bosons,$G_{1},G_{2}$, and four mass 
eigenstates, $A_{1},A_{2},A_{3},A_{4}$.

\section{Single charged fields in MSUSY331.}

On this case the basis is given by 
$\eta_{1}^{+}, \rho^{+}, \eta_{1}^{\prime +}, \rho^{\prime +},
\eta_{2}^{+}, \chi^{+},\eta_{2}^{\prime +},\chi^{\prime +}$, with the help of 
Eq.(\ref{constraintequationsinmodel1}), we get
\begin{eqnarray}
{\bf M_{15}}&=&{\bf M_{16}}={\bf M_{17}}={\bf M_{18}}=0 \,\ , \nonumber \\
{\bf M_{25}}&=&{\bf M_{26}}={\bf M_{27}}={\bf M_{28}}=0 \,\ , \nonumber \\
{\bf M_{35}}&=&{\bf M_{36}}={\bf M_{37}}={\bf M_{38}}=0 \,\ , \nonumber \\
{\bf M_{45}}&=&{\bf M_{46}}={\bf M_{47}}={\bf M_{48}}=0 \,\ , \nonumber \\
{\bf M_{11}}&=&- \frac{f_{1}}{18}v^{2}_{\rho}+ \frac{g^{2}}{4}
(v^{\prime 2}_{\eta}+v^{2}_{\rho}-v^{\prime 2}_{\rho})-  \frac{\sqrt{2}}{12}
(6k_{1}\frac{v_{\rho}v_{\chi}}{v_{\eta}}+f_{1}^{\prime}
\frac{v^{\prime}_{\rho}v^{\prime}_{\chi} \mu_{\eta}}{v_{\eta}}) \nonumber \\ 
&+& \frac{\sqrt{2}f_{1}}{12v_{\eta}}(v^{\prime}_{\rho}\mu_{\rho}v_{\chi}+
v_{\rho}\mu_{\chi}v^{\prime}_{\chi}) \,\ , \nonumber \\
{\bf M_{12}}&=&- \frac{f^{2}_{1}}{18}v_{\eta}v_{\rho}+ \frac{g^{2}}{4} 
v_{\eta}v_{\rho}- \frac{k_{1}}{\sqrt{2}}v_{\chi}+ \frac{f_{1}}{6 \sqrt{2}}
\mu_{\chi}v^{\prime}_{\chi} \,\ , \nonumber \\
{\bf M_{13}}&=&- \frac{g^{2}}{4} v_{\eta}v^{\prime}_{\eta} \,\ , \nonumber \\ 
{\bf M_{14}}&=&- \frac{g^{2}}{4} v_{\eta}v^{\prime}_{\rho}+ 
\frac{f_{1}}{6 \sqrt{2}}\mu_{\rho}v_{\chi}- \frac{f^{\prime}_{1}}{6 \sqrt{2}}
\mu_{\eta}v^{\prime}_{\chi} \,\ , \nonumber \\
{\bf M_{22}}&=& \frac{f_{1}^{2}}{18}v^{2}_{\eta}+ \frac{\sqrt{2}f_{1}}{12} 
\left( \frac{\mu_{\chi}v_{\eta}v^{\prime}_{\chi}}{v_{\rho}}- 
\frac{\mu_{\eta}v^{\prime}_{\eta}v_{\chi}}{v_{\rho}} \right) + 
\frac{g^{2}}{4}(v^{2}_{\eta}-v^{\prime 2}_{\eta}+v^{\prime 2}_{\rho})- 
\frac{\sqrt{2}k_{1}}{6}\frac{v_{\eta}v_{\chi}}{v_{\rho}} \nonumber \\ &+& 
\frac{\sqrt{2}f^{\prime}_{1}}{36}
\frac{\mu_{\rho}v^{\prime}_{\eta}v^{\prime}_{\chi}}{v_{\rho}} \,\ , 
\nonumber \\
{\bf M_{23}}&=&- \frac{g^{2}}{4}v^{\prime}_{\eta}v_{\rho}- 
\frac{f_{1}}{6 \sqrt{2}}\mu_{\eta}v_{\chi}+ \frac{f^{\prime}_{1}}{6 \sqrt{2}}
\mu_{\rho}v^{\prime}_{\chi} \,\ , \nonumber \\
{\bf M_{24}}&=&- \frac{g^{2}}{4}v_{\rho}v^{\prime}_{\rho} \,\ , \nonumber \\ 
{\bf M_{33}}&=&- \frac{f^{\prime 2}_{1}}{18}v^{\prime}_{\rho}+ \frac{g^{2}}{4}
(v^{2}_{\eta}-v^{2}_{\rho}-v^{\prime 2}_{\rho})+ 
\frac{\sqrt{2}f^{\prime}_{1}}{12} \left( 
\frac{\mu_{\chi}v^{\prime}_{\rho}v_{\chi}}{v^{\prime}_{\eta}}+
\frac{\mu_{\rho}v_{\rho}v^{\prime}_{\chi}}{v^{\prime}_{\eta}} \right) - 
\frac{\sqrt{2}f_{1}}{12} \frac{\mu_{\eta}v_{\rho}v_{\chi}}{v^{\prime}_{\eta}} 
\nonumber \\ &-& 
\frac{k_{1}}{\sqrt{2}}
\frac{\mu_{\eta}v^{\prime}_{\rho}v^{\prime}_{\chi}}{v^{\prime}_{\eta}} 
\,\ , \nonumber \\ 
{\bf M_{34}}&=&- \frac{f^{\prime 2}_{1}}{18}v^{\prime}_{\eta}v^{\prime}_{\rho}+
\frac{g^{2}}{4}v^{\prime}_{\eta}v^{\prime}_{\rho}+ 
\frac{f^{\prime}_{1}}{6\sqrt{2}}\mu_{\chi}v_{\chi}- 
\frac{k^{\prime}_{1}}{\sqrt{2}}v^{\prime}_{\chi} \,\ , \nonumber \\
{\bf M_{44}}&=&- \frac{f^{\prime 2}_{1}}{18}v^{\prime 2}_{\eta}+ 
\frac{g^{2}}{4}(v^{\prime 2}_{\eta}-v^{2}_{\eta}+v^{2}_{\rho})+
\frac{\sqrt{2}f_{1}}{12}\frac{\mu_{\rho}v_{\eta}v_{\chi}}{v^{\prime}_{\rho}}-
\frac{k^{\prime}_{1}}{\sqrt{2}}
\frac{v^{\prime}_{\eta}v^{\prime}_{\chi}}{v^{\prime}_{\rho}}+
\frac{\sqrt{2}f^{\prime}_{1}}{12}
\frac{\mu_{\chi}v^{\prime}_{\eta}v_{\chi}}{v^{\prime}_{\rho}} \nonumber \\ &-&
\frac{\sqrt{2}f^{\prime}_{1}}{12}
\frac{\mu_{\eta}v_{\eta}v^{\prime}_{\chi}}{v^{\prime}_{\rho}}
\,\ , \nonumber \\ 
{\bf M_{55}}&=&- \frac{k_{1}}{\sqrt{2}}\frac{v_{\rho}v_{\chi}}{v_{\eta}}+
\frac{\sqrt{2}f_{1}}{12}\frac{\mu_{\rho}v^{\prime}_{\rho}v_{\chi}}{v_{\eta}}-
\frac{f^{2}_{1}}{12}v^{2}_{\chi}+\frac{\sqrt{2}f^{\prime}_{1}}{12}
\frac{\mu_{\eta}v^{\prime}_{\rho}v^{\prime}_{\chi}}{v_{\eta}}+
\frac{\sqrt{2}f_{1}}{12}\frac{\mu_{\chi}v_{\rho}v^{\prime}_{\chi}}{v_{\eta}} 
\nonumber \\ &+&
\frac{g^{2}}{4}(v^{\prime 2}_{\eta}+v^{2}_{\chi}-v^{\prime 2}_{\chi}) \,\ , 
\nonumber \\
{\bf M_{56}}&=&- \frac{k_{1}}{\sqrt{2}}v_{\rho}+ \frac{f_{1}}{6 \sqrt{2}}
\mu_{\rho}v^{\prime}_{\rho}- \frac{f^{2}_{1}}{18}v_{\eta}v_{\chi}+
\frac{g^{2}}{4}v_{\eta}v_{\chi} \,\ , \nonumber \\
{\bf M_{57}}&=&- \frac{g^{2}}{4}v_{\eta}v^{\prime}_{\eta} \,\ , \nonumber \\ 
{\bf M_{58}}&=&- \frac{f^{\prime}_{1}}{6 \sqrt{2}}\mu_{\eta}v^{\prime}_{\rho}+
\frac{f_{1}}{6 \sqrt{2}}\mu_{\chi}v_{\rho}-
\frac{g^{2}}{4}v_{\eta}v^{\prime}_{\chi} \,\ , \nonumber \\
{\bf M_{66}}&=&- \frac{k_{1}}{\sqrt{2}}\frac{v_{\eta}v_{\rho}}{v_{\chi}}+
\frac{\sqrt{2}f_{1}}{12}\frac{\mu_{\rho}v^{\prime}_{\rho}v_{\eta}}{v_{\chi}}-
\frac{\sqrt{2}f_{1}}{12}\frac{\mu_{\eta}v_{\rho}v^{\prime}_{\eta}}{v_{\chi}}+
\frac{\sqrt{2}f^{\prime}_{1}}{12}
\frac{\mu_{\chi}v^{\prime}_{\rho}v^{\prime}_{\eta}}{v_{\chi}}-
\frac{f^{2}_{1}}{12}v^{2}_{\eta} \nonumber \\ &+&
\frac{g^{2}}{4}(v^{\prime 2}_{\eta}+v^{\prime 2}_{\chi}-v^{\prime 2}_{\eta}) 
\,\ , \nonumber \\ 
{\bf M_{67}}&=&- \frac{f_{1}}{6 \sqrt{2}}\mu_{\eta}v_{\rho}+
\frac{f^{\prime}_{1}}{6 \sqrt{2}}\mu_{\chi}v^{\prime}_{\rho}-
\frac{g^{2}}{4}v_{\chi}v^{\prime}_{\eta} \,\ , \nonumber \\
{\bf M_{68}}&=&- \frac{g^{2}}{4}v_{\chi}v^{\prime}_{\chi} \,\ , \nonumber \\
{\bf M_{77}}&=&- \frac{k_{1}}{\sqrt{2}}
\frac{v^{\prime}_{\rho}v^{\prime}_{\chi}}{v^{\prime}_{\eta}}+
\frac{\sqrt{2}f_{1}}{12}\frac{\mu_{\eta}v_{\rho}v_{\chi}}{v^{\prime}_{\eta}}-
\frac{\sqrt{2}f_{1}}{12}
\frac{\mu_{\chi}v^{\prime}_{\rho}v_{\chi}}{v^{\prime}_{\eta}}+
\frac{\sqrt{2}f_{1}}{12}
\frac{\mu_{\rho}v^{\prime}_{\chi}v_{\rho}}{v^{\prime}_{\eta}}-
\frac{\sqrt{2}f^{\prime}_{1}}{12}
\frac{\mu_{\chi}v^{\prime}_{\rho}v^{\prime}_{\eta}}{v_{\chi}} \nonumber \\ &-&
\frac{f^{2}_{1}}{12}v^{\prime 2}_{\chi}+
\frac{g^{2}}{4}(v^{\prime 2}_{\eta}-v^{2}_{\chi}-v^{\prime 2}_{\chi}) \,\ , 
\nonumber \\ 
{\bf M_{78}}&=&- \frac{k^{\prime}_{1}}{\sqrt{2}}v^{\prime}_{\rho}+ 
\frac{f^{\prime}_{1}}{6 \sqrt{2}}\mu_{\rho}v_{\rho}- 
\frac{f^{\prime 2}_{1}}{18}v^{\prime}_{\eta}v^{\prime}_{\chi}+
\frac{g^{2}}{4}v^{\prime}_{\eta}v^{\prime}_{\chi} \,\ , \nonumber \\
{\bf M_{88}}&=&- \frac{k^{\prime}_{1}}{\sqrt{2}}
\frac{v^{\prime}_{\rho}v^{\prime}_{\eta}}{v^{\prime}_{\chi}}+ \frac{g^{2}}{4}
(v^{2}_{\chi}+v^{\prime 2}_{\eta}-v^{2}_{\eta})+  
\frac{\sqrt{2}f^{\prime}_{1}}{12} \left( 
\frac{\mu_{\rho}v^{\prime}_{\eta}v_{\rho}}{v^{\prime}_{\chi}}- 
\frac{\mu_{\eta}v_{\eta}v^{\prime}_{\rho}}{v^{\prime}_{\chi}} \right)
- \frac{\sqrt{2}f_{1}^{\prime}}{12}
\frac{\mu_{\chi}v_{\eta}v_{\rho}}{v^{\prime}_{\chi}} \nonumber \\
&-& \frac{f^{\prime 2}_{1}}{12}v^{\prime 2}_{\eta}.
\label{matchar1m1} 
\end{eqnarray}

Analysing the matrix elements above, we conclude that the mixing occurs  in 
the set $\eta_{1}^{+},\rho^{+},\eta_{1}^{\prime +},\rho^{\prime +}$ and in the 
set of $\eta_{2}^{+},\chi^{+},\eta_{2}^{\prime +},\chi^{\prime +}$ and it is 
in agreement with the results presented in \cite{tonasse}.

This results mean that in the most general case, the singly charged scalars 
are obtained by diagonalizing one $8 \times 8$ matrix. Ignoring the null 
elements mixing, the matrix presented above decompose into two series 
of $4 \times 4$ matrices. 

The mixing between the particles $\eta_{1}^{+},\rho^{+},\eta_{1}^{\prime +},
\rho^{\prime +}$ is giving by the following matrix
\begin{equation}
\left( \begin{array}{cccc}
{\bf M_{11}} & {\bf M_{12}} & {\bf M_{13}} & {\bf M_{14}} \\
{\bf M_{21}} & {\bf M_{22}} & {\bf M_{23}} & {\bf M_{24}} \\
{\bf M_{31}} & {\bf M_{32}} & {\bf M_{33}} & {\bf M_{34}} \\
{\bf M_{41}} & {\bf M_{42}} & {\bf M_{43}} & {\bf M_{44}}
\end{array} \right) \,\ ,
\label{mixeta1rho}
\end{equation}
using the values given at Eq.(\ref{matchar1m1}) we get one Goldstone boson, 
$G^{+}_{1}$, and three mass eigenstates $H^{+}_{1},H^{+}_{2},H^{+}_{3}$. While 
the second mixing is given by
\begin{equation}
\left( \begin{array}{cccc}
{\bf M_{55}} & {\bf M_{56}} & {\bf M_{57}} & {\bf M_{58}} \\
{\bf M_{65}} & {\bf M_{66}} & {\bf M_{67}} & {\bf M_{68}} \\
{\bf M_{75}} & {\bf M_{76}} & {\bf M_{77}} & {\bf M_{78}} \\
{\bf M_{85}} & {\bf M_{86}} & {\bf M_{87}} & {\bf M_{88}}
\end{array} \right) \,\ .
\label{mixeta2chi}
\end{equation}
This matrix, with the help of Eq.(\ref{matchar1m1}), has one Goldstone, 
$G^{+}_{2}$, three mass eigenstates $H^{+}_{4},H^{+}_{5},H^{+}_{6}$. 

\section{Double charged fields in MSUSY331.}

On this case, with the help of Eq.(\ref{constraintequationsinmodel1}), our 
results are
\begin{eqnarray}
{\bf {\cal M}_{11}}&=&- \frac{k_{1}}{\sqrt{2}} 
\frac{v_{\eta}v_{\chi}}{v_{\rho}}- \frac{\sqrt{2}f_{1}}{12}\left(
\frac{\mu_{\eta}v^{\prime}_{\eta}v_{\chi}}{v_{\rho}}-
\frac{\mu_{\rho}v^{\prime}_{\eta}v_{\chi}}{v_{\rho}}- 
\frac{\mu_{\chi}v_{\eta}v^{\prime}_{\chi}}{v_{\rho}} \right) -
\frac{f^{2}_{1}}{12}v^{2}_{\chi}+ \frac{g^{2}}{4}(v^{\prime 2}_{\rho}+
v^{2}_{\chi}-v^{\prime}_{\chi}) \,\ , \nonumber \\ 
{\bf {\cal M}_{12}}&=&- \frac{k_{1}}{\sqrt{2}}v_{\eta}- 
\frac{f_{1}}{6 \sqrt{2}}\mu_{\eta}v^{\prime}_{\eta}- \frac{f^{2}_{1}}{18}
v_{\rho}v_{\chi}+ \frac{g^{2}}{4}v_{\rho}v_{\chi} \,\ , \nonumber \\
{\bf {\cal M}_{13}}&=&- \frac{g^{2}}{4}v_{\rho}v^{\prime}_{\rho} \,\ , 
\nonumber \\
{\bf {\cal M}_{14}}&=&\frac{f^{\prime}_{1}}{6 \sqrt{2}}\mu_{\rho}
v^{\prime}_{\eta}+ \frac{f_{1}}{6 \sqrt{2}}\mu_{\chi}v_{\eta}- 
\frac{g^{2}}{4}v_{\rho}v^{\prime}_{\chi} \,\ , \nonumber \\
{\bf {\cal M}_{22}}&=&- \frac{k_{1}}{\sqrt{2}} 
\frac{v_{\eta}v_{\rho}}{v_{\chi}}+ \frac{\sqrt{2}f_{1}}{12}
(\mu_{\rho}v^{\prime}_{\rho}v_{\eta}-\mu_{\eta}v^{\prime}_{\eta}v_{\rho})+
\frac{\sqrt{2}f^{\prime}_{1}}{12}
\frac{\mu_{\chi}v^{\prime}_{\eta}v^{\prime}_{\rho}}{v_{\chi}}- 
\frac{f^{2}_{1}}{13}v^{2}_{\rho}+ \frac{g^{2}}{6}(v^{2}_{\rho}+
v^{\prime 2}_{\rho}+v^{\prime 2}_{\chi}), \nonumber \\ 
{\bf {\cal M}_{23}}&=&\frac{f^{\prime}_{1}}{6 \sqrt{2}}\mu_{\chi}
v^{\prime}_{\eta}+ \frac{f_{1}}{6 \sqrt{2}}\mu_{\rho}v_{\eta}- 
\frac{g^{2}}{4}v^{\prime}_{\rho}v_{\chi} \,\ , 
\nonumber \\
{\bf {\cal M}_{24}}&=&- \frac{g^{2}}{4}v_{\chi}v^{\prime}_{\chi} \,\ , 
\nonumber \\
{\bf {\cal M}_{33}}&=&- \frac{k^{\prime}_{1}}{\sqrt{2}} 
\frac{v^{\prime}_{\eta}v^{\prime}_{\chi}}{v^{\prime}_{\rho}}+
 \frac{\sqrt{2}f_{1}}{12}
\frac{\mu_{\rho}v_{\chi}v_{\eta}}{v^{\prime}_{\rho}}+
\frac{\sqrt{2}f^{\prime}_{1}}{12v^{\prime}_{\rho}}
\frac{\mu_{\chi}v^{\prime}_{\eta}v_{\chi}}{v_{\chi}}+ 
\frac{f^{\prime}_{1}}{12v^{\prime}_{\rho}}\mu_{\eta}v_{\eta}v^{\prime}_{\chi}+ 
\frac{g^{2}}{6}(v^{2}_{\rho}+v^{\prime 2}_{\chi}-v^{2}_{\chi}) \,\ , 
\nonumber \\
{\bf {\cal M}_{34}}&=&- \frac{k^{\prime}_{1}}{\sqrt{2}}v^{\prime}_{\eta}- 
\frac{f^{\prime}_{1}}{6 \sqrt{2}}\mu_{\eta}v_{\eta}- 
\frac{f^{\prime 2}_{1}}{18}v^{\prime}_{\rho}v^{\prime}_{\chi}+ 
\frac{g^{2}}{4}v^{\prime}_{\rho}v^{\prime}_{\chi} \,\ , \nonumber \\
{\bf {\cal M}_{44}}&=&- \frac{k^{\prime}_{1}}{\sqrt{2}} 
\frac{v^{\prime}_{\eta}v^{\prime}_{\rho}}{v^{\prime}_{\chi}}+
 \frac{\sqrt{2}f_{1}}{12} \left(
\frac{\mu_{\rho}v_{\rho}v^{\prime}_{\eta}}{v^{\prime}_{\chi}}-
\frac{\mu_{\chi}v_{\eta}v^{\prime}_{\rho}}{v^{\prime}_{\chi}} \right) +
\frac{\sqrt{2}f_{1}}{12} \frac{\mu_{\chi}v_{\eta}v_{\rho}}{v^{\prime}_{\chi}}-
\frac{f^{\prime 2}_{1}}{13} v^{\prime 2}_{\rho} \nonumber \\ &+&
\frac{g^{2}}{6}(v^{2}_{\rho}+v^{\prime 2}_{\rho}+v^{2}_{\chi}).
\label{matdcharm1} 
\end{eqnarray}
This mass matrix has one Goldstone boson, $G^{++}_{1}$, and three mass 
eigenstates $H^{++}_{1},H^{++}_{2},H^{++}_{3}$.

\end{document}